\documentclass[12pt]{article}

\usepackage{latexsym} 
\usepackage{amssymb}  
\usepackage{amsbsy}   
\usepackage{graphicx}       
\usepackage[dvips]{color}

\newcommand{\De}{\Delta}

\newcommand{\si}{\sigma}

\newcommand{\Om}{\Omega}

\newcommand{\imp}{~\Rightarrow}
\newcommand{\p}{\partial}
 
\newcommand{\txt}{\textstyle}

\newcommand{\dsp}{\displaystyle}

\newcommand\eqn[1]{(\ref{#1})}      
\newcommand\Eqn[1]{Eq.~(\ref{#1})}  

\newcommand{\beq}{\begin{equation}}
\newcommand{\eeq}{\end{equation}}
\newcommand{\ba}{\begin{array}}
\newcommand{\bea}{\begin{eqnarray}}
\newcommand{\ea}{\end{array}}
\newcommand{\eea}{\end{eqnarray}}
\newcommand{\bi}{\begin{itemize}}  
\newcommand{\ei}{\end{itemize}}
\newcommand{\ben}{\begin{enumerate}} 
\newcommand{\een}{\end{enumerate}}
\newcommand{\bc}{\begin{center}}
\newcommand{\ec}{\end{center}}
\newcommand{\bl}{\begin{flushleft}}
\newcommand{\el}{\end{flushleft}}
\newcommand{\br}{\begin{flushright}}
\newcommand{\er}{\end{flushright}}

\newcommand\comment[1]{ \hbox{[{\it Comment suppressed here.}\/]} }
\newcommand\hide[1]{}

\renewcommand{\O}{{\cal O}}

\newcommand{\ie}{{i.e.}}

\newcommand{\skipover}[1]{}

\newcommand{\third}{{\txt \frac{1}{3}}}

\newcommand{\twothirds}{{\txt \frac{2}{3}}}

\makeatletter 

\def\appendix{\par                              
    \setcounter{section}{0}                     
    \setcounter{subsection}{0}
    \renewcommand{\theequation}{\Alph{section}.\arabic{equation}}
    \renewcommand{\thesection}{Appendix \Alph{section}
                \setcounter{equation}{0}  } 
}

\def\applabel#1{\@bsphack
  \protected@write\@auxout{}%
         {\string\newlabel{#1}{{\Alph{section}}{\thepage}}}%
  \@esphack}

\def\section{
\setcounter{equation}{0}        
\@startsection {section}{1}{\z@}{-3.5ex plus -1ex minus 
 -.2ex}{2.3ex plus .2ex}{\large\bf}}
\renewcommand{\theequation}{\arabic{section}.\arabic{equation}}

\def\subsection{\@startsection{subsection}{2}{\z@}{-3.25ex plus -1ex minus 
 -.2ex}{1.5ex plus .2ex}{\normalsize\bf}}

\def\subsubsection{\@startsection{subsubsection}{3}{\z@}{-3.25ex plus
 -1ex minus -.2ex}{1.5ex plus .2ex}{\normalsize}}

\makeatother   

\newcommand{\fm}{{\rm fm}} 
\newcommand{\km}{{\rm km}} 
 
\newcommand{\MeV}{{\rm MeV}}

\newcommand{\Msolar}{M_\odot} 
\newcommand{\Mmax}{M_{\rm max}} 
\newcommand{\nsat}{n_{\rm sat}} 
\newcommand{\rhocrit}{\rho_{\rm c}}

\newcommand{\nova}[1]{#1}

\begin{document}

\title{\bf Hybrid stars that masquerade as neutron stars}

\author{
Mark Alford${}^{(a)}$, Matt Braby${}^{(a)}$,
Mark Paris${}^{(b,c)}$, Sanjay Reddy${}^{(c)}$ 
\\[0.5ex]
\parbox{0.6\textwidth}%
{
\normalsize
\begin{itemize}
\item[${}^{(a)}$] 
  Physics Department, \\
  Washington University, \\
  St.~Louis, MO 63130, U.S.A.
\item[${}^{(b)}$] 
  Jefferson National Laboratory, \\
  12000 Jefferson Avenue \\
  Newport News, VA 23606
\item[${}^{(c)}$] 
  Theoretical Division, \\
  Los Alamos National Laboratory, \\
  Los Alamos, NM~87545, U.S.A.
\end{itemize}
}}

\newcommand{\preprintno}{
  \normalsize LA-UR-04-8091
}

\date{March 15, 2005 \\[1ex] \preprintno}

\begin{titlepage}
\maketitle
\renewcommand{\thepage}{}          

\begin{abstract}
We show that a hybrid (nuclear + quark matter) star can have
a mass-radius relationship very similar to
that predicted for a star made of purely nucleonic matter.
We show this for a generic parameterization
of the quark matter equation of state, and also for an MIT bag model,
each including a phenomenological correction based on gluonic
corrections to the equation of state. We obtain
hybrid stars as heavy as $2~\Msolar$
for reasonable values of the bag model parameters.
For nuclear matter, we use the equation of state 
calculated by Akmal, Pandharipande, and Ravenhall using many-body
techniques. Both mixed and homogeneous phases of nuclear and 
quark matter are considered.
\end{abstract}

\end{titlepage}

\renewcommand{\thepage}{\arabic{page}}

\section{Introduction}
\label{sec:intro}

It has long been hypothesized that some compact stars might 
actually be ``hybrid stars'' containing
cores of quark matter. The observationally accessible features of
compact stars include their mass and radius, and there have been
various investigations of how the presence of a quark matter core
would affect the mass-radius relationship of a compact star.
The general conclusion has been that quark matter softens the equation
of state, so that hybrid stars are predicted to have lower maximum mass 
($M\lesssim 1.7~\Msolar$) than nuclear matter\footnote{
In this paper we only consider two-flavor nuclear matter.
Introducing hyperons or kaon condensation into the nuclear matter
also softens the equation of state and lowers the maximum mass.
\cite{Lattimer:2000nx}.
} stars
($M\lesssim 2.2~\Msolar$) 
\cite{Lattimer:2000nx,Maieron:2004af,AlfRed,Buballa:2003et,Banik:2002kc,Burgio_etal,Gocke:2001ri,Schertler:2000xq}.
On the observational side, all mass measurements are currently
compatible, at the $2\si$ level, with $M\lesssim 1.7~\Msolar$
\cite{Thorsett:1999,Lattimer:2004pg}.  However, some are near the
limit of compatibility, for example pulsar J0751+1807 in a white
dwarf-neutron star binary system, whose mass is currently measured at
$2.1(4)~\Msolar$ \cite{Nice:2004fn}.  The error bars on these
measurements will decrease over time so quark matter cores may seem to
be on the point of being ruled out.  Our purpose in this paper is to
show that in fact it will be harder to rule out quark matter via
$M(R)$ observations than these simple considerations indicate.

In an earlier paper \cite{AlfRed}, we performed mass-radius
calculations for hybrid and pure quark matter stars using a simple MIT
bag model equation of state. In that model, correlations arising due
to quark-quark interactions were neglected. However, the effects of the
strange quark mass and corrections to the equation of state due to the
pairing energy associated with color
superconductivity\cite{colorSCreviews}, the formation of quark
Cooper pairs, were incorporated.  We found that color
superconductivity boosts the pressure of the quark matter relative to
nuclear matter, lowering the transition density (at fixed bag
constant), but that the maximum mass was similar to that obtained in
other work, namely about $1.6\Msolar$.  We found that in order to form
stars near this upper bound it was necessary to set the bag constant
to a low value, so that quark matter is very nearly stable, and the
NM$\to$QM phase transition occurs below nuclear saturation density,
$\nsat=0.16 \mbox{ fm}^{-3}$: the heaviest stars consisted almost
entirely of quark matter, with only a thin crust of nuclear matter.

\subsection{Overview of this study}
In this paper we will again use equations of state based on the MIT
bag model
for the quark matter, but we will include an additional parameter that
imitates the effect of including perturbative QCD corrections. 

For the first stage of this analysis (Sec.~\ref{sec:stage1})
we will actually not use any model at all. We write down a purely
phenomenological quark matter equation of state, 
consisting simply of a power series
expansion in the quark chemical potential, $\mu$
\beq
\Omega_{\rm QM} = - \frac{3}{4\pi^2}~a_4~\mu^4 + \frac{3}{4\pi^2}
a_2~\mu^2 + B_{\rm eff},
\label{QM_EoS}
\eeq
where $a_4,a_2$ and $B_{\rm eff}$ are independent of $\mu$.
We will show that for 
$a_4\approx 0.7$ (which we will see later is physically reasonable), 
one can obtain heavy hybrid stars ($M\approx 2~\Msolar$)
while still ensuring that the NM$\to$QM phase transition occurs
above nuclear saturation density. In fact, we show that it is possible
to mimic the mass-radius behavior of nucleonic stars over a wide
range of masses. 

We then go on to the second stage of our analysis, in which we use
a quark matter equation of state based on
a physical model: competition between a ``normal'' unpaired
quark matter phase and the CFL color-superconducting phase, in
an MIT-bag-model formalism.
This corresponds to
giving $a_2$ and $B_{\rm eff}$ a simple step-function dependence 
on the chemical potential, with the step occurring at the 
transition between these phases. The details are determined by the
microscopic parameters of the model, the strange quark mass $m_s$ and
pairing gap $\De$.

The coefficient $a_4\equiv 1-c$ is a rough parameterization of
QCD corrections to the pressure of the free-quark Fermi sea,
and previous calculations \cite{Fraga} show that the value $a_4\approx 0.7,
c\approx 0.3$ is reasonable.
The physical model also allows us to calculate the behavior of
charged phases, and hence to study inhomogeneous ``mixed'' phases 
of nuclear matter and quark matter.

The results from the quark matter model turn out to be very similar to 
those obtained
with the simple parameterization.
We also find that, given our current (and, likely,
future) ignorance of the high-density values of basic parameters like
the strange quark mass and bag constant, there are no characteristic
features of the $M(R)$ relationship that could be used to verify the
presence of color superconductivity in the quark matter core.

Our conclusion is that the maximum mass of hybrid stars is mainly determined
by the size of the QCD corrections to the coefficient $a_4$ in the
quark matter equation of state \eqn{QM_EoS}, and that for reasonable
values of $a_4$ hybrid stars can be as heavy as $2~\Msolar$.
This result is robust:
it is not affected when we move up to a more sophisticated model, nor by the
introduction of mixed phases. There is therefore little
reason to expect it to change when the model is made even more complicated,
e.g.~by including kaon-condensed \cite{kaons}, crystalline \cite{glitch}, 
mixed \cite{Reddy:2004my}
or gapless \cite{g2SC,gCFL} phases, or
allowing continuous $\mu$-dependence in the strange quark mass
or pairing gap (see below).

\subsection{Other approaches}
Our calculations use a basic MIT bag model, in which the bag
constant and quark masses are assumed to be density-independent. Other
approaches are certainly possible.  Within the MIT bag model one can
use a density-dependent bag constant, although this does not
appreciably change the maximum mass prediction \cite{Burgio_etal}.
The density-dependence of the constituent quark masses and the color
superconducting gap $\De$ can be estimated by using an NJL model
instead of a bag model, with coupled mean-field Schwinger-Dyson
equations for the masses and gaps.  Such models give a high effective
bag constant and quark masses, and typically predict small numbers of
strange quarks, with 2SC rather than CFL color superconductivity, but
again the maximum masses turn out to be of order $1.6~\Msolar$
\cite{Buballa:2003et,Gocke:2001ri}, although folding
certain Gaussian form
factors into the 4-fermion interaction can give masses up to
$1.8~\Msolar$ \cite{Grigorian:2003vi}. While these models are well
motivated theoretically, their specific predictions relating to the
density dependencies of quark masses and the effective bag constant
remain untested. In this work we adopt a minimal approach and retain
quartic and quadratic powers of the chemical potential in our
expression for the free energy. This will allow us to do a parameter
study independent of any specific model. It would be interesting to
see whether including a reasonable estimate of QCD corrections in the
NJL model increases the maximum mass in that context also.

\section{The nuclear matter equation of state}
\label{sec:NM_EoS}
Our treatment of nuclear matter is completely standard: at densities 
above half nuclear
saturation density ($\nsat$)
we employ the equation of state of Akmal, Pandharipande, and Ravenhall
(APR) \cite{APR98}. At lower densities we use the
standard tabulated low-density equation of state \cite{BPS,NV}. Our
previous
studies of hybrid stars \cite{AlfRed,ARRW} used the relativistic mean
field Walecka model \cite{SW} to describe nuclear matter or APR for
beta stable charge-neutral nuclear matter. The relativistic mean field
is an effective description of the nuclear matter that is constrained
by properties of nuclear matter at saturation density. The mean field
approximation ignores many-body correlations that could play an
important role. APR use the variational chain summation (VCS) to
include these correlations in calculating the equation of state of
nucleon matter \cite{PW79,AP97}. They employ a realistic
non-relativistic Hamiltonian with the Argonne $v_{18}$ \cite{WSS95}
two-body potential and the Urbana IX \cite{PPCW95} three nucleon
interaction. An equation of state as a function of baryon density and
proton fraction, $x_p$ is obtained by interpolating between the pure
neutron matter (PNM) $x_p=0$ and symmetric nuclear matter (SNM)
$x_p=0.5$ results using a generalized Skyrme interaction containing
momentum and density dependent delta function interactions \cite{PR89}
described below.

The energy is evaluated for a variational wave function which
takes into account many-body correlation effects. It is 
composed as a symmetrized product of two-body correlation operators,
$F_{ij}$ acting on the Fermi gas wave function. The correlation operators 
are written as a sum of terms which include operators appearing in the 
Hamiltonian. The two-body cluster contribution to the energy is 
minimized by $F_{ij}$ which satisfy the Euler-Lagrange equations 
determined within this ansatz. Heuristically, two-body
operators which appear in the Hamiltonian induce correlations between
particles whose spatial dependence is approximated by solving a two-body
Schr\"{o}dinger-like equation subject to suitable boundary conditions.
The variation of the wave function is effected at the two-body level
by varying parameters appearing in this equation. Though this wave
function neglects three-body correlations it is estimated to be accurate 
to a few MeV/nucleon in SNM and about one MeV/nucleon in PNM 
at nuclear density. This
accuracy is achieved through the inclusion of many-body effects via the
VCS technique. Comparable accuracy is obtained with the 
other many-body techniques like the Brueckner-Bethe-Goldstone 
method of Ref.\cite{Baldo:1997}.

The APR equation of state exhibits a transition from a low density
phase (LDP) to a high density phase (HDP) having spin-isospin order,
possibly due to neutral pion condensation, in PNM at a density of
$\sim 0.20$ fm$^{-3}$ and in SNM at $\sim 0.32$ fm$^{-3}$. The VCS
calculations of the energy of PNM and SNM are extrapolated to general
values of $x_p$ using a function of the form \bea \epsilon_N(\rho,x_p)
&=& \left(\frac{\hbar^2}{2m_N}+f(\rho,x_p)\right)\tau_p
+\left(\frac{\hbar^2}{2m_N}+f(\rho,1-x_p)\right)\tau_n \nonumber \\
&+&g(\rho,x_p=0.5)(1-(1-2x_p)^2)+g(\rho,x_p=0)(1-2x_p)^2 \eea
motivated by a generalized Skyrme interaction.  Here $\epsilon_N$ is
the total nuclear energy density, $\tau_{n,p}$ are the neutron and
proton Fermi gas kinetic densities, and $f(\rho,x_p)$ and
$f(\rho,1-x_p)$ are functions which parameterize the effective mass of
the nucleons and $g(\rho,x_p=0)$ and $g(\rho,x_p=0.5)$ are potential
energy terms. These functions are parameterized to fit the energies of
PNM and SNM calculated in VCS. Separate parameterizations are used in
the LDP and HDP for the functions $g$. The energy density of nuclear
matter will be used to determine the allowed equation of state of
hybrid stars for the case of a sharp transition to quark matter and to
determine the allowed phases for a mixed transition to quark matter.

\section{A simple phenomenological quark matter equation of state}
\label{sec:stage1}

One of the main points of this paper is to show the effects of
including a parameter in the quark matter equation of state that
roughly incorporates the effects of
gluon-mediated QCD interactions between the quarks in the Fermi sea.
We first do this in the context of a simple parameterization of
the quark matter equation of state, and show that the resultant
hybrid stars can have mass-radius relations very similar to that of
pure nuclear stars, with mass up to $2~\Msolar$. Later
(Sec.~\ref{sec:stage2}) we will show that these conclusions
remain true in a more sophisticated model.

The simple phenomenological parameterization of
the quark matter equation of state is
\beq
\ba{rcl}
\Omega_{\rm QM} &=&\dsp - \frac{3}{4\pi^2}~a_4~\mu^4 + \frac{3}{4\pi^2}
a_2~\mu^2 + B_{\rm eff}\ , \\[2ex]
a_4 &\equiv& 1-c \ ,
\ea
\label{QM_EoS1}
\eeq
where the parameters $a_4,a_2,B_{\rm eff}$ are independent of $\mu$.

\medskip\noindent 1) \underline{The quartic coefficient $a_4=1-c$}.\\
For quark matter consisting of 3 flavors of
free non-interacting quarks, $c=0$, so $a_4=1$ (see
the discussion of the physical model of quark matter in
Sec.~\ref{sec:stage2}).
However, once QCD corrections are taken into account, we expect $c\neq 0$. 
The QCD corrections to the quark matter equation of state
were first evaluated to $\O(\alpha_s^2)$ by Freedman and McLerran
\cite{pertEoS}; then Fraga, Pisarski and Schaffner-Bielich
\cite{Fraga} (FPS) showed that the
$\O(\alpha_s^2)$ pressure for three massless flavors
can be approximated by a bag-model-inspired form,
\begin{equation}
P_{\alpha_s^2} \approx
\dsp \frac{3}{4\pi^2}~(1-c)~\mu^4 - B_{\rm eff}\ .\\[2ex]
\label{peff}
\end{equation}
Matching to the $\O(\alpha_s^2)$ perturbative calculations in the interval 
$\mu\simeq 300-600~\MeV$, they found that
$B_{\rm eff}$ varies widely with renormalization scale
(see also Ref.~\cite{Andersen:2002jz}), but they
consistently find $c\approx 0.37$.
We do not use FPS's specific values of
$B_{\rm eff}$ and $c$ because, as they observe, the QCD coupling is
strong at the density of interest for compact star physics, so there
is no reason to expect the leading order calculation to be
accurate. However, we take their results as indicating that 
QCD corrections are not negligible, so one should
include $c$ as an additional parameter in the quark matter
equation of state, with a value of order $0.3$.

\medskip\noindent 2) \underline{The quadratic coefficient $a_2$}.\\ As
we will see in Sec.~\ref{sec:stage2}, the $\mu^2$ term can arise
from the strange quark mass (which increases the free energy) or color
superconductivity (which reduces it). If chiral symmetry remains
broken in the light quark sector due to a robust $\langle
\bar{q}q\rangle$ condensate, then the large ($\sim 100-300$ MeV)
constituent quark masses of the up and down quarks would also result
in $\mu^2$ term similar to that due to the strange quark mass.
For color-flavor-locked quark matter,
$a_2 = m_s^2-4 \De^2$.  For now, we simply include $a_2$ as a
phenomenological parameter.

\medskip\noindent 3) \underline{The bag constant and the 
  transition density $\rhocrit$}.\\
In our parameterization the effective bag constant simply accounts for
the free energy contribution that is independent of $\mu$. While this
is related to the vacuum pressure its numerical value in our
parameterization need not be the same as in early bag-model studies of hadron
phenomenology \cite{Chodos:1974pn}. The effective bag constant is
unknown and difficult to calculate or measure, so we will not use it
as a parameter, since that would obscure the fact that part of the
effect of varying other parameters may simply be a renormalization of
the unknown parameter $B_{\rm eff}$.  To expose the physically significant
effects of varying $a_4, a_2, B_{\rm eff}$ we will specify a more physical
quantity, the maximum density $\rhocrit$ of nuclear matter, i.e.~the
density at which the NM$\to$QM transition occurs.  The structure of
the star is then calculated as a function of $a_4, a_2$ and $\rhocrit$.  

Two subtle points arise in such a reparameterization.  
Firstly, as we will see below, for values of the perturbative
correction parameter $c$ around 0.3, the quark matter and APR nuclear
matter equations of state have almost exactly the same shape over a
wide range of pressures. This can lead to multiple phase transitions
back and forth between NM and QM. Of course, when the two phases have almost
identical equations of state, it does not matter (for mass and radius 
calculations) where transitions between them occur. We therefore
simply choose
$\rhocrit$ to be associated with the location of the first transition.

Secondly, when we use the more complicated model in Sec.~\ref{sec:stage2},
we will allow for the possibility of mixed phases, which
blur out the NM$\to$QM transition over a range of densities and
pressures, making it hard to identify ``the'' transition density.
However, for the purpose of fixing the bag constant we do not
have to allow mixed phases. We
will therefore define $B_{\rm eff}(a_4,a_2,\rhocrit)$ as the value 
of the bag constant that
would give a sharp NM$\to$QM transition at a nuclear matter density
$\rhocrit$ if only charge neutral bulk phases were permitted
(as would happen if the
NM-QM surface tension were $\geqslant 40~\MeV/\fm^3$).

\subsection{The physical effects of the ``perturbative correction'' parameter}
\label{sec:c}

We now discuss the physical importance of the QCD correction $c$.
Firstly, it is clear analytically that, at fixed bag
constant, $c$ has very little effect on the relation $E(p)$ between
energy density and pressure for quark matter,
which enters into the 
Tolman Oppenheimer Volkoff (TOV) equation \cite{TOV}. This can be seen
by setting $a_2=0$ (i.e.~neglecting quark masses and pairing), in which case
\beq
\ba{rcl}
p &=& \dsp \phantom{3}(1-c)\frac{3 \mu^4}{4\pi^2}  - B_{\rm eff} \\[2ex]
E &=& \dsp 3(1-c)\frac{3 \mu^4}{4\pi^2} + B_{\rm eff} \\[2ex]
\imp E &=& 3p+4B_{\rm eff} \qquad\mbox{independent of $c$}
\ea
\eeq
However, this does not mean that $c$ is unimportant.
Clearly $c$ makes a dramatic difference to $p(\mu)$, so
it strongly affects the position of the NM$\to$QM transition.
Moreover, as described above, we are not working at fixed $B_{\rm eff}$:
when we change $c$ we keep the transition density $\rhocrit$ fixed,
with a resultant change in $B_{\rm eff}$.

\begin{figure}[tb]
\begin{center}
\includegraphics[width=0.47\textwidth]{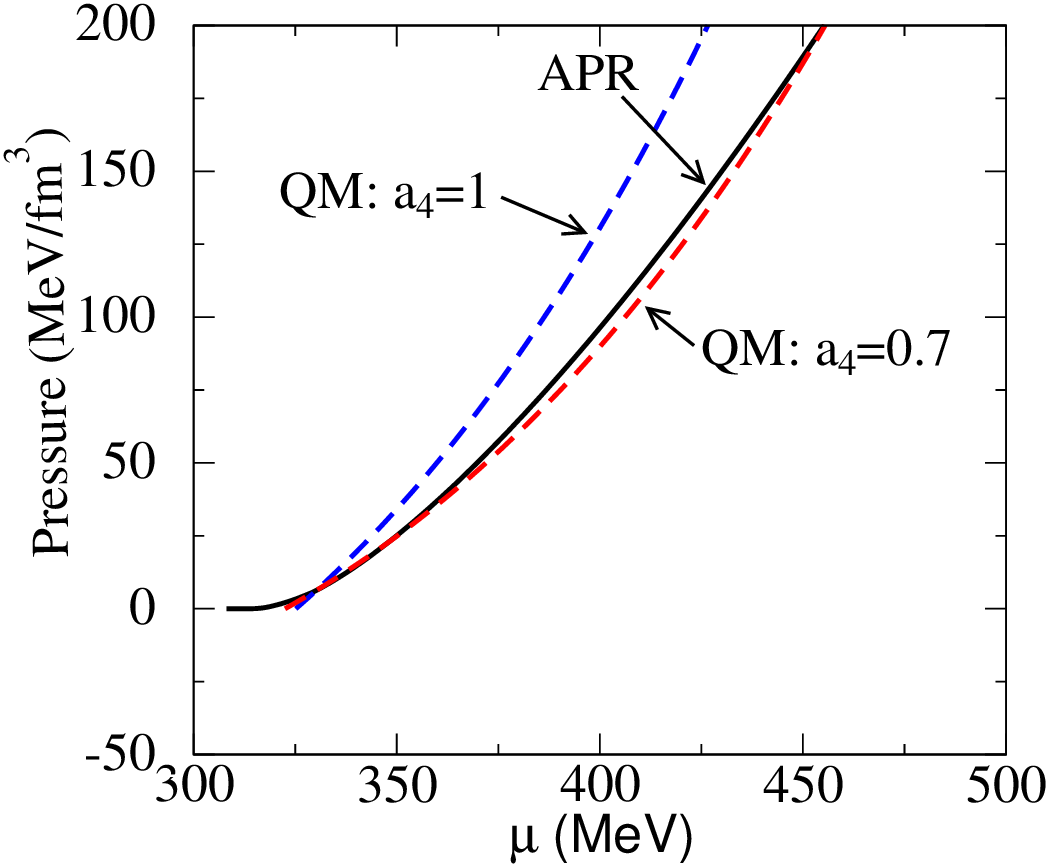}
\hspace{0.02\textwidth}
\includegraphics[width=0.47\textwidth]{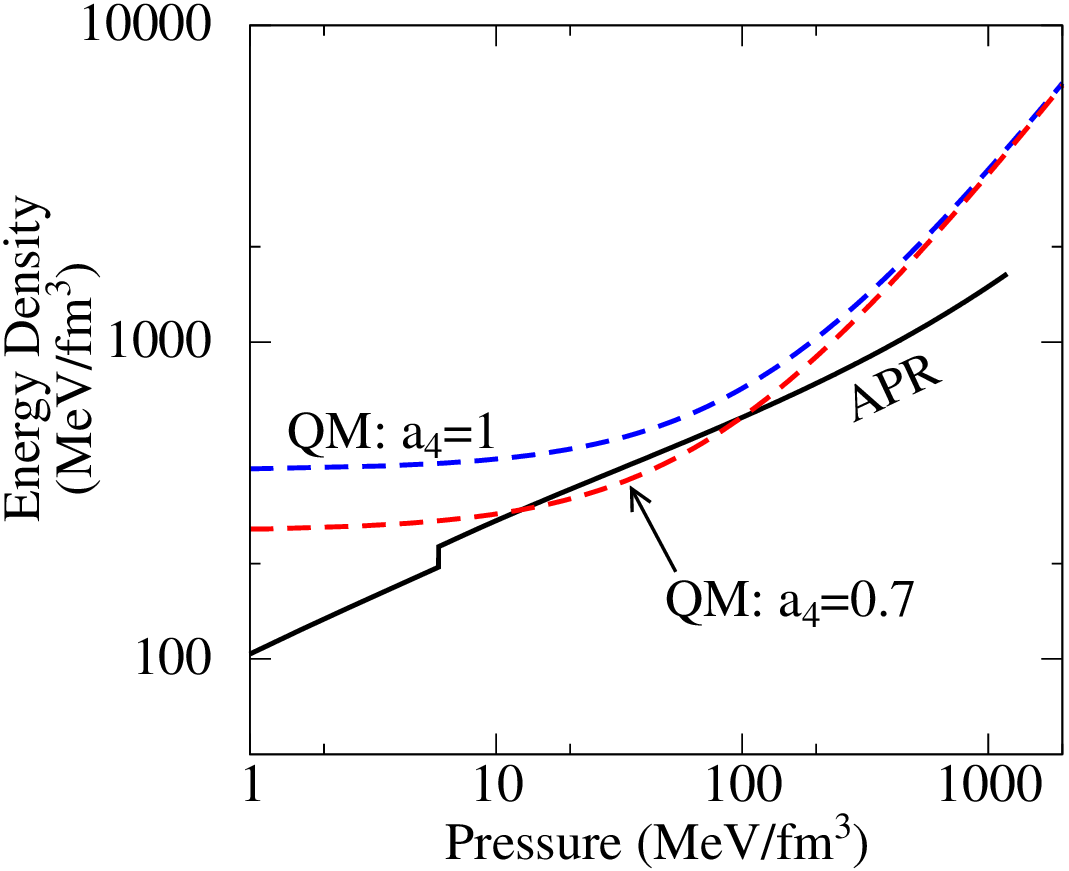}
\end{center}
\caption{\small
Equations of state for APR nuclear matter (solid line) and
for quark matter with the phenomenologically
parameterized equation of state \eqn{QM_EoS1} with $a_2=(150~\MeV)^2$
(dashed curves: upper curve has no perturbative correction 
($c=0$), lower one has $c=0.3$). 
For each quark matter equation of state the bag constant
$B_{\rm eff}$ was fixed by requiring that nuclear matter give way to
quark matter at $\rhocrit=1.5\,\nsat$.
The figure shows that when we include perturbative-type
corrections, quark matter has a $p(\mu)$ relation
almost identical to that of APR nuclear matter, 
and its $E(p)$ is also very
similar over the range of pressures that is relevant to
compact star masses.
}
\label{fig:EoS_fixrho}
\end{figure}

\begin{figure}[tb]
\begin{center}
\includegraphics[width=0.46\textwidth,angle=0]{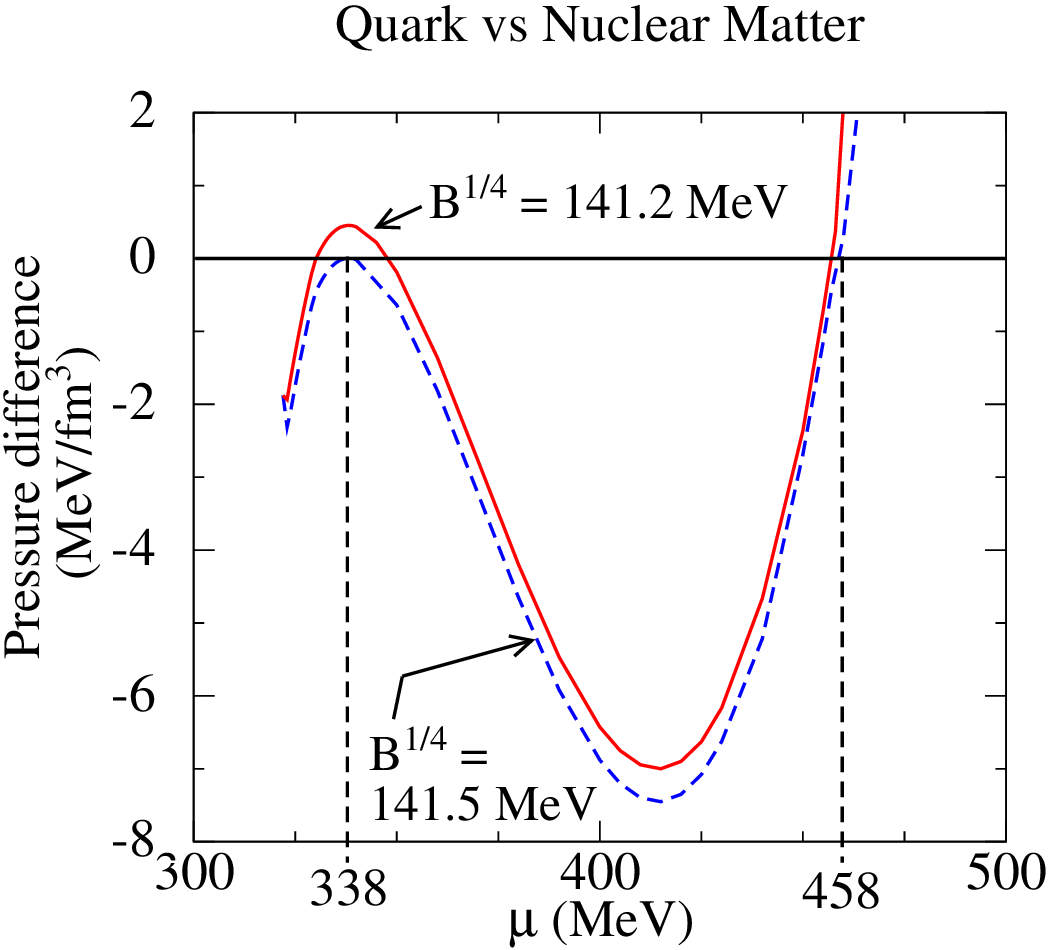}
\includegraphics[width=0.42\textwidth,angle=0]{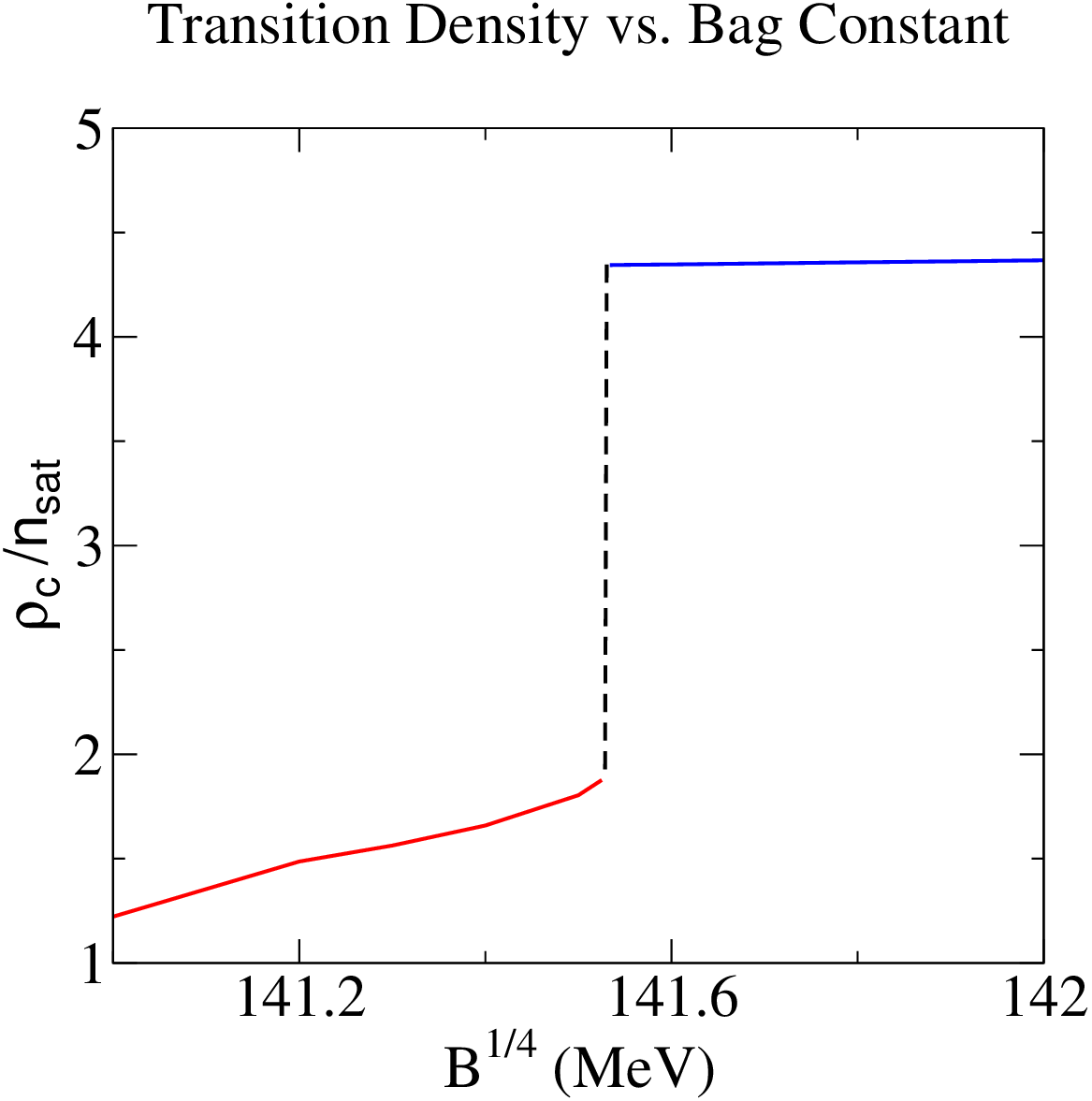}
\end{center}
\caption{\small The pressure difference between quark matter and
APR nuclear matter for $a_4=0.7$, $a_2=(150~\MeV)^2$, 
illustrating how the near-identity of
their $p(\mu)$ relations leads to difficulties in defining a unique
transition density. In the left panel,
the solid curve is for $B_{\rm eff}^{1/4}=141.2~\MeV$ and shows three
transitions, the first occurring at $\mu=324.4~\MeV$, corresponding
to $\rhocrit=1.5\,\nsat$ (Fig.~\ref{fig:EoS_fixrho}). Raising $B_{\rm
eff}$ gives a transition at higher $\mu$, but it is technically
impossible to achieve a (first) transition in the range $338<\mu<458~\MeV$
($1.9\,\nsat<\rho<4.3\,\nsat$): 
at $B_{\rm eff}^{1/4}=141.5~\MeV$ (dashed curve)
the transition jumps from the bottom of this range to the top. The
right panel illustrates how this translates to a
discontinuous behavior in dependence of
$\rhocrit$ on $B_{\rm eff}$.}
\label{fig:transitions}
\end{figure}

We illustrate the effect of non-zero $c$ in Fig.~\ref{fig:EoS_fixrho},
which shows $p(\mu)$ and $E(\mu)$ for APR nuclear matter
and for the phenomenological
description of quark matter, with $a_2=(150~\MeV)^2$,
for the cases $c=0$ and $0.3$
(tuning $B_{\rm eff}$ to keep the transition density $\rho=1.5\,\nsat$).
We see that the $c=0.3$ quark matter equation of state is very similar to APR
over the pressure range $10$ to $200~\MeV/\fm^3$. In fact, close examination
of the $p(\mu)$ relationship shows that there are three phase transitions
with increasing density: from APR nuclear matter at low density
to quark matter, then back to APR, then back to quark matter again.
On this basis, we expect stars containing quark matter with $c\approx 0.3$
to show $M(R)$ curves very similar to those of nuclear matter stars,
making them correspondingly difficult to rule out from $M(R)$ measurements
alone. In Sec.~\ref{sec:maxM} we will see that this is indeed the case.

The $p(\mu)$ relations for $c=0.3$ quark matter and APR are so similar that it
can be difficult to say where the phase transition really occurs.  In
Fig.~\ref{fig:transitions} we show the pressure difference for the
parameters of Fig.~\ref{fig:EoS_fixrho} ($B_{\rm eff}^{1/4}=141.2~\MeV$, solid
curve) and for a higher value of the bag constant
($B_{\rm eff}^{1/4}=141.5~\MeV$, dashed curve).  
At $B_{\rm eff}^{1/4}=141.2~\MeV$, the
three transitions are clearly visible.  We see that when the bag
constant is raised to $B_{\rm eff}^{1/4}=141.5~\MeV$ to obtain a transition at
higher density, the value of the chemical potential at the transition,
and hence the density, jumps discontinuously from $\mu=338~\MeV$
($\rho=1.9\,\nsat$) to $\mu=458~\MeV$ ($\rho=4.3\,\nsat$). This is
illustrated in the right panel of
Fig.~\ref{fig:transitions}.  It is therefore technically impossible to
choose $B_{\rm eff}$ so as to obtain a (first) transition density in the range
$1.9\,\nsat<\rhocrit<4.3\,\nsat$.

It is very interesting to speculate on the possibility of multiple
phase transitions inside compact stars, but this feature is highly
sensitive to the precise relative shapes of our 
quark matter and APR equations
of state. We cannot claim to know these to the level of accuracy (a
few percent, i.e.~a few $\MeV/\fm^3$) that would be required to say
whether the number of transitions is 1, 3, or even 5. Our main message
in this paper is that for reasonable values of the quark matter
parameters the equations of state may be very similar. A realistic
interpretation of Fig.~\ref{fig:transitions} is that at
$B_{\rm eff}^{1/4}=141.5~\MeV$ the critical density jumps rapidly 
from $1.9\,\nsat$ to $4.3\,\nsat$. 
Any $\rhocrit$ in that range therefore corresponds to
$B_{\rm eff}^{1/4}=141.5~\MeV$. 
(Pictorially, this corresponds to blurring out the
curves in Fig.~\ref{fig:transitions} by a few $\MeV/\fm^3$.) This is
the procedure we will follow in determining $B_{\rm eff}$ for given
$\rhocrit,m_s,\De$ when we investigate how the maximum compact star mass
depends on these parameters.

\nova{
One can extrapolate from Figs.~\ref{fig:EoS_fixrho} and
\ref{fig:transitions} to predict what will happen for even larger
values of $c$.  For suitable values of the bag constant there will be
a small range of densities (near nuclear density) where quark matter
has higher pressure than APR nuclear matter, then at intermediate
densities APR will be favored again, and finally quark matter will win
at high densities. When such equations of state are used to construct
compact stars, the result will be a star containing a shell of quark
matter with APR nuclear matter outside and inside it. This also leads to
large masses, but only because most of the star is nuclear matter.  In
the rest of this paper we will discuss values
of $c$ up to $0.3$, which is suggested by Fraga et.~al.'s
fit to the two-loop equation of state \cite{Fraga}, and also 
offers the possibility of heavy stars with a considerable quark
matter fraction.
}

\subsection{Mass-radius relationship: a hybrid star that ``looks nuclear''}
\label{sec:MR}

\begin{figure}[tb]
\begin{center}
 \includegraphics[width=0.8\textwidth,angle=0]{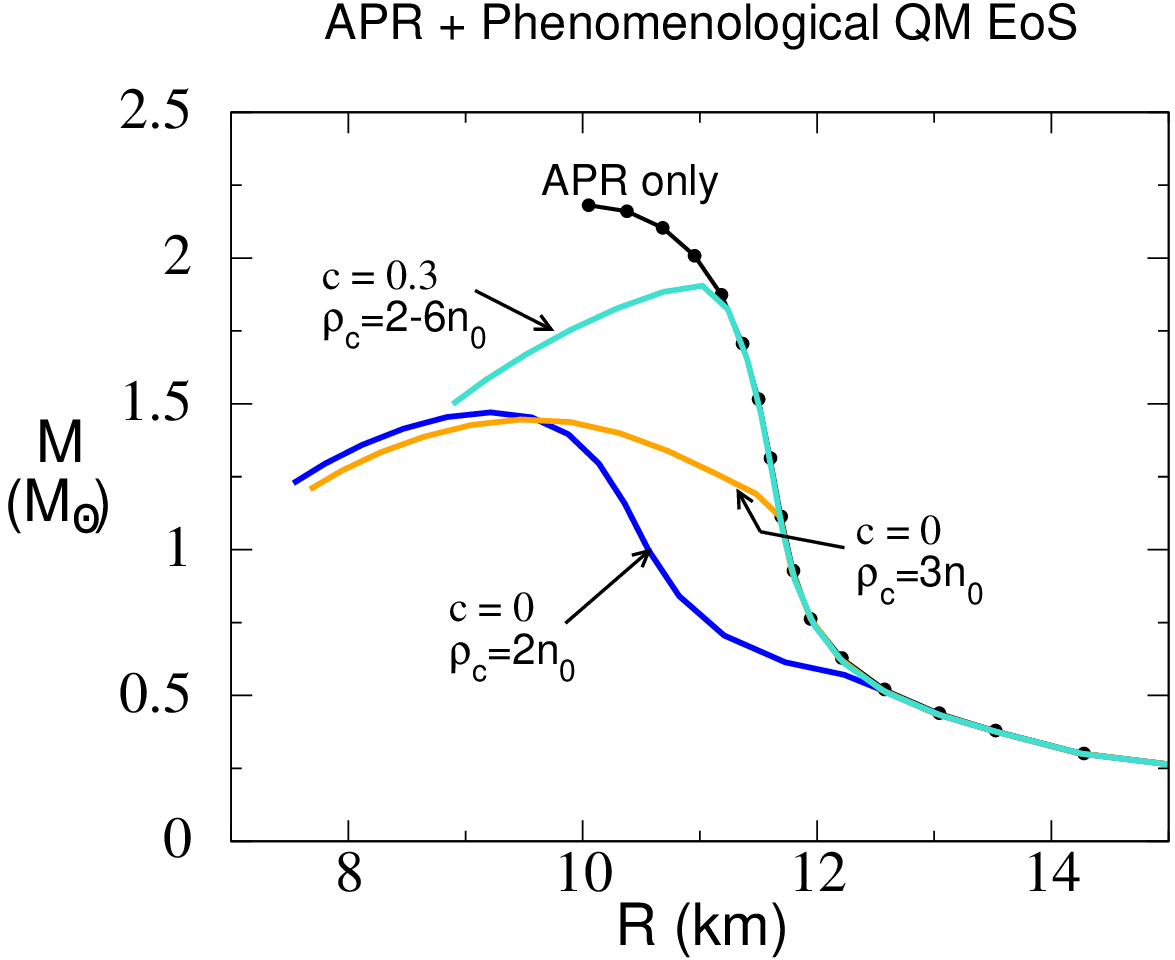}
\end{center}
\caption{\small
$M(R)$ relationship for hybrid stars involving quark matter 
obeying the
phenomenological equation of state \eqn{QM_EoS1} with
$a_2=(180~\MeV)^2$. In each case the bag constant was chosen
to give the desired transition density.
The line with black dots is the $M(R)$ relation for a pure nuclear
APR star. All the phases are neutral and homogeneous.
We see that the maximum mass $\Mmax$ is very sensitive to the 
QCD correction $c$, but not to the transition density.
For $c=0.3$, which is close to the value suggested
in Ref.~\cite{Fraga}, $\Mmax\approx 1.9~\Msolar$.
Note also that the $c=0.3$ equation of state is so similar to APR that
its $M(R)$ curve is almost identical to that of pure APR.
}
\label{fig:stage1_heavystar}
\end{figure}

\begin{figure}[tb]
\begin{center}
 \includegraphics[width=0.8\textwidth,angle=0]{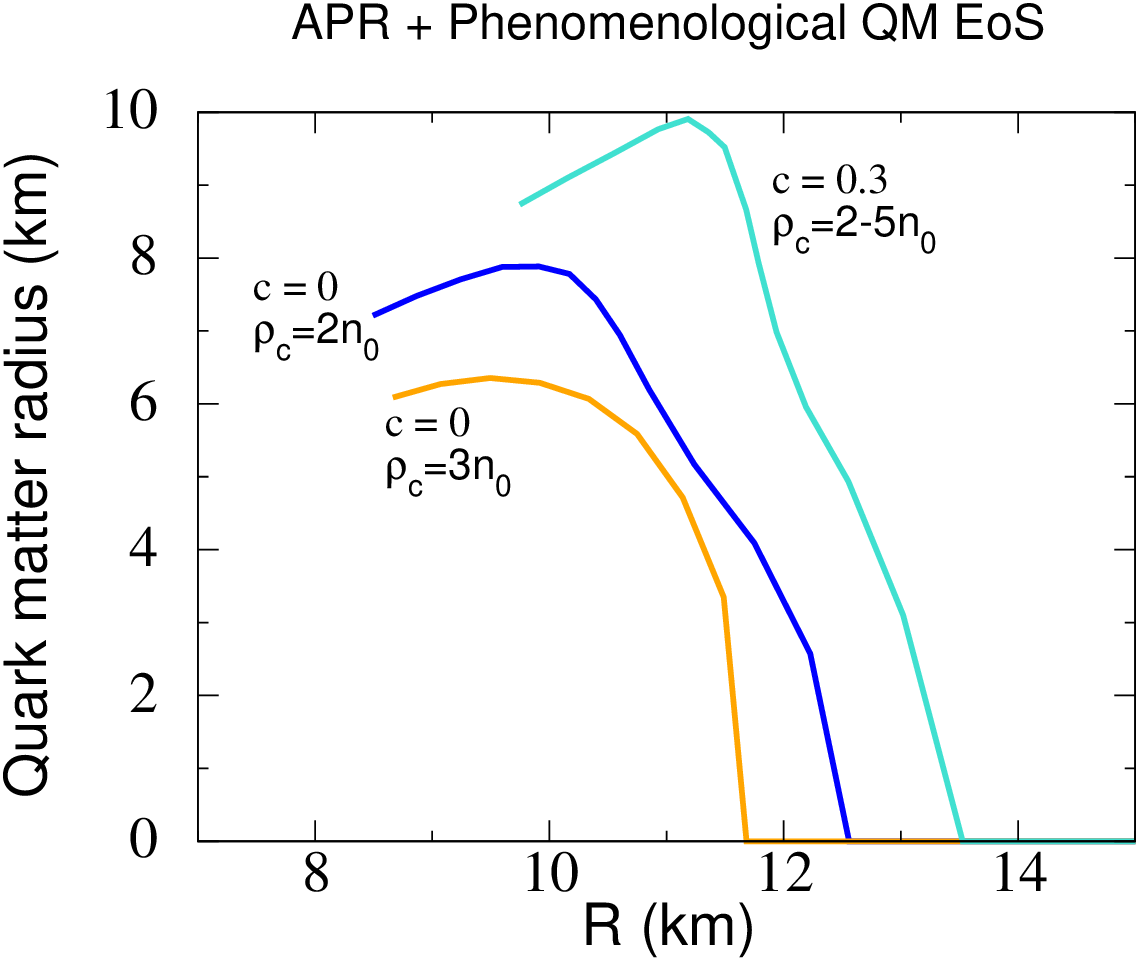}
\end{center}
\caption{\small
\nova{
The radius at which the first (lowest-density) transition 
from nuclear matter to quark matter occurs, for the families of stars
whose $M(R)$ relations were given in Fig.~\ref{fig:stage1_heavystar}.
Note that this radius is quite large for the heavier stars.
In the $c=0.3$ case there may be transitions back and forth between
quark and nuclear matter at higher density (smaller radii).
}
}
\label{fig:stage1_quarkradius}
\end{figure}

Before surveying a wide range of values of the parameters $a_4=1-c,\,
a_2,\, \rhocrit$, we first display the $M(R)$ curve for some specific
cases, showing how masses approaching $2~\Msolar$ can be achieved.
The $M(R)$ relation is obtained by solving the TOV equation in the
standard way, as described in Ref.~\cite{AlfRed}. We choose a range of
central pressures, and integrate the TOV equation outward until the
pressure drops to zero, which marks the surface of the star, and the
integrated energy density yields the mass \cite{GBOOK}.  In
Fig.~\ref{fig:stage1_heavystar} we show the resulting $M(R)$ curves for quark
matter with $a_2=(180~\MeV)^2$. We performed calculations with
$c=0$ and with $c=0.3$ \cite{Fraga}. We tuned the bag
constant to give a range of homogeneous neutral-matter transition
densities $\rhocrit=2$ to $4\,\nsat$.

The most obvious feature is the dependence on $c$: compared to the
stars with $c=0$,
hybrid stars containing quark matter with larger perturbative
correction $c=0.3$ are significantly heavier and a little larger,
with masses approaching $2~\Msolar$, and radii around $11~\km$.
It is also striking that stars with $c=0.3$
have an $M(R)$ relation essentially identical to that of APR nuclear
matter, up to masses around $1.9~\Msolar$. This is because
quark matter with $c=0.3$ has an equation of state
very close to that of APR nuclear matter over the relevant range of
pressures. This was discussed in Sec.~\ref{sec:c}. 

\nova{
In Fig.~\ref{fig:stage1_quarkradius} we show, for the same family of stars,
the radius at which the first (lowest-density) transition 
from nuclear matter to quark matter occurs.
Note that this radius is quite large for the heavier stars.
For values of $c$ close to 0.3 
there are transitions back and forth between
quark and nuclear matter as one goes deeper into the star, because
the equations of state are so similar.
While such multiple transitions cannot be ruled out a priori, we suspect 
that they are not physical. They are an
artifact of our model description
of the nuclear and quark phases and not a robust prediction. 
Here and in the rest of the paper
we will ignore this possibility and entertain only one transition
from nuclear to quark matter, beyond which we will use the quark matter 
equation of state.  
While this will make little difference to the structure of the star,
since the nuclear and quark matter equations of state are both very
similar in this regime, it would have important consequences for
transport properties.
}

\subsection{Maximum mass as a function of quark matter parameters}
\label{sec:maxM}

Having seen that with a perturbative correction $c$ set to a reasonable
value we can increase the maximum hybrid star mass,
we now look at how maximum hybrid star mass depends on
the parameters of our phenomenological
equation of state \eqn{QM_EoS1}: 
$a_4$, $a_2$, and the transition density $\rhocrit$.

\begin{figure}[tb]
\begin{center}
\includegraphics[width=0.8\textwidth,angle=0]{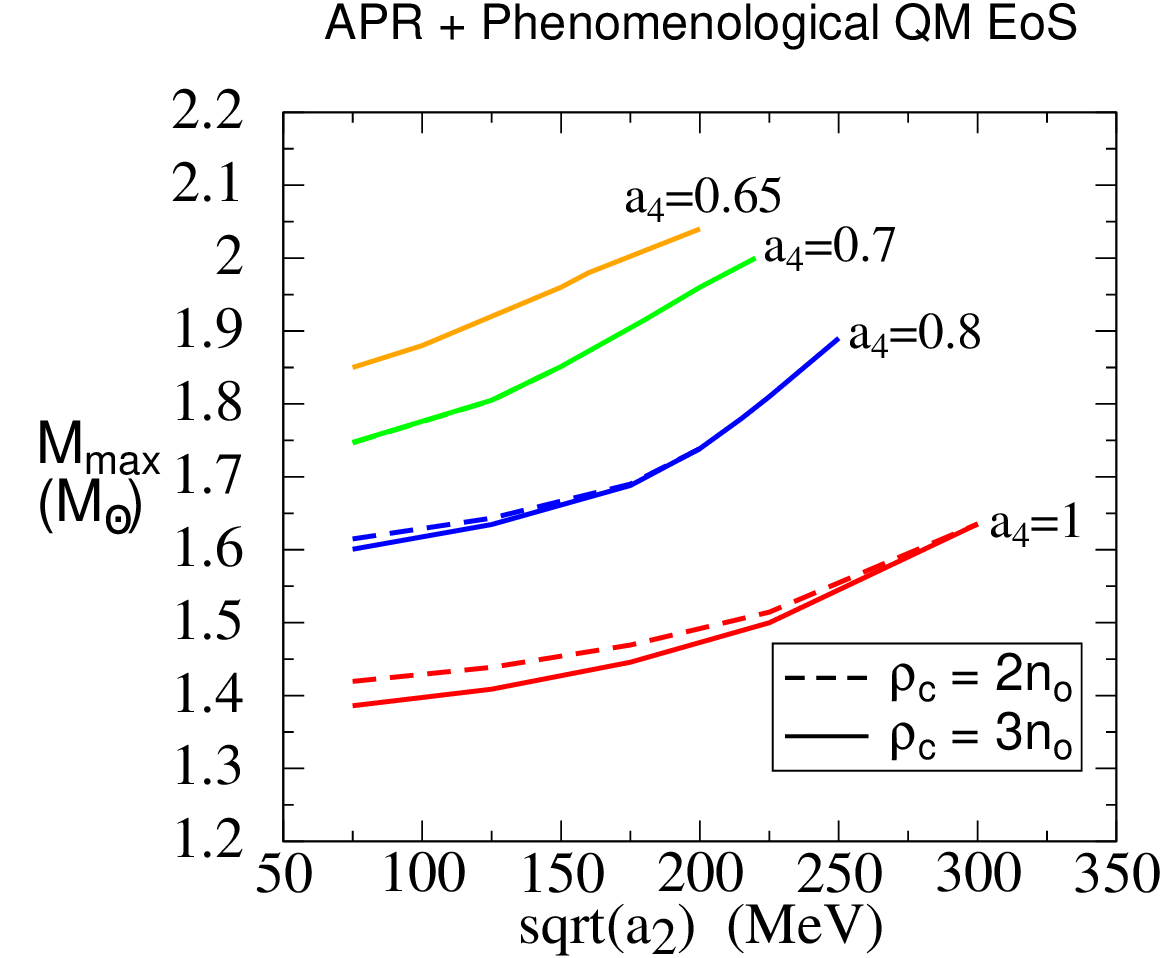}
\end{center}
\caption{\small
Dependence of $\Mmax$ on $a_2$ for hybrid stars involving quark matter 
obeying the phenomenological equation of state \eqn{QM_EoS1}.
Increasing $a_2$ decreases $B_{\rm eff}$ at fixed $\rhocrit$, 
giving rise to larger maximum masses. 
We show results for two
transition densities, $\rhocrit=3\,\nsat$ and $\rhocrit=2\,\nsat$.
The maximum mass is not very sensitive to the transition density.
}
\label{fig:stage1_vary_y}
\end{figure}

In Fig.~\ref{fig:stage1_vary_y} we plot the maximum mass, $\Mmax$
obtained by varying the central pressure, and choosing
the heaviest resulting stable star that contains some
quark matter.
We calculated $\Mmax$ as a function of $a_2$, for
$a_4=1,\ 0.8,\ 0.7,\ 0.65$. In each case we repeated the
calculation for two different transition densities,
$\rhocrit=3\,\nsat$ (solid lines) and $\rhocrit=2\,\nsat$ (dashed lines).
The lines in Fig.~\ref{fig:stage1_vary_y} end when there is no
longer a stable hybrid star. This corresponds
to the hybrid star branch in the $M(R)$ plot having a positive slope
along its whole length.

We see that masses up to $2~\Msolar$ can be obtained
by choosing a small value of $a_4$, and setting
$\sqrt{a_2}$ to an appropriate value
in the range $150~\MeV$ to $250~\MeV$. 
(Recall that in unpaired quark matter $\sqrt{a_2}$
corresponds to the strange quark mass.)
This is understandable since the
effective bag constant must decrease with increasing $a_2$ at fixed
$\rhocrit$: it is well known that in stars dominated by quark matter the
$\Mmax \sim 1/\sqrt{B_{\rm eff}}$ \cite{Witten:1984rs}.

One might want to ask how large the quark matter core is for the
heaviest stars, but this question does not have an easy answer.
For $a_4\sim 0.7$ ($c\sim 0.3$),
the quark matter and nuclear matter equations of state are so
similar that the transition density is not precisely defined (as we
saw in Fig.~\ref{fig:transitions}) so it is not clear where the
transition occurs: there is a whole range of transition densities that
correspond to the same bag constant.  This is why, for lower values of
$a_4$, the maximum mass is not very sensitive to the transition
density. 
\nova{
Fig.~\ref{fig:stage1_quarkradius} shows that the first
transition from nuclear to quark matter occurs at quite large radii
for the heavy stars.
}

\subsection{Conclusion}
We have seen that the very simple phenomenological parameterization
\eqn{QM_EoS1}
of the quark matter equation of state allows us to survey the 
main effects of the various parameters, and makes it clear that
$a_4=1-c$ is the most important in determining the maximum mass.
With a value corresponding to free quark matter ($c=0$) we 
find a maximum hybrid star mass
of about $1.6~\Msolar$, similar to the value found in previous studies.
With a value corresponding to
QCD corrections of a plausible strength ($c\approx 0.35$) we
find a maximum mass of about $2~\Msolar$.

In the next section we show that this conclusion is robust.
We repeat our analysis with a more complicated quark matter equation
of state, based on the expected physics of dense quark matter.
We allow competition between paired and unpaired phases and mixed 
quark and nuclear matter phases. We find that these factors have no 
effect on our essential findings regarding the maximum mass of hybrid stars.

\section{A more sophisticated approach: a physical model of quark matter,
and mixed phases}
\label{sec:stage2}

The model we will use in this section 
is more complicated and physically well-founded than the
phenomenological parameterization of Sec.~\ref{sec:stage1}, but it 
is still a relatively simple MIT bag model.
We will allow competition between two phases of quark matter, the 
three-flavor unpaired
phase and the color-flavor-locked (CFL) color-superconducting phase.
The phase with the lowest free energy (highest pressure) is favored.

The parameters of this model are the strange quark mass $m_s$, the
CFL pairing gap $\De$, the density $\rhocrit$ of the nuclear to
quark matter transition (which determines the bag constant),
and a QCD-inspired correction parameter $c$. This is similar to
the phenomenological parameterization of Sec.~\ref{sec:stage1},
but the effective parameter $a_2$ is now calculated within a model.

At asymptotically high chemical potential on the order of $10^8$ MeV, the 
CFL phase is known to be the ground state of 3-flavor quark matter, though
it may persist to much lower chemical potential.
At lower densities the strange quark mass becomes non-negligible
relative to the chemical potential, and at
$m_s^2/\mu = 4\De$ there is a transition to unpaired quark matter \cite{AR-02}.
\footnote{Actually, before that, at $m_s^2 = 2\mu\De$, there is a transition
to a gapless phase \cite{gCFL}, and at
$m_s \sim m_{\rm light}^{1/3}\Delta^{2/3}$ there is the
possibility of $K^0$ condensation in the CFL phase
\cite{kaons}:
we ignore these additional complications because their contributions
to the pressure are of order $m_s^4$, which is just a renormalization
of the bag constant, which for us is a non-physical parameter.}
When we include the nuclear matter, there is a three-way competition
between nuclear matter, unpaired quark matter, and CFL quark matter.

Now that we have a physical model in hand, 
we can calculate the pressure of charged as well as neutral phases
of quark matter. This means that we
can study mixed as well as homogeneous phases. The transition from
nuclear matter to quark matter can proceed via a mixed phase
 \cite{Glendenning:1992vb}, in which there is 
charge separation,
and a positively charged nuclear phase interpenetrates with a negatively 
charged quark matter phase, yielding a globally neutral
inhomogeneous phase. This will only occur if the surface tension
at the boundary of the two phases is low enough. Otherwise
there will be a sharp interface between the two homogeneous neutral
phases.
The critical
surface tension is $\si_c\approx 40~\MeV/\fm^3$
\cite{Heiselberg:1992dx,ARRW}.  
Since the surface tension is
completely unknown, we will separately consider both mixed phases
and sharp interfaces in our calculations.

\subsection{Unpaired 3-flavor quark matter}

In the naive bag model, where perturbative corrections are ignored, the
free energy of free quark matter consists of the kinetic contribution from
a degenerate free gas of three colors of relativistic quarks and the 
negative vacuum pressure from the bag constant, $B_{\rm eff}$
\beq
\Omega_{\mathrm{unp}}(\mu_u,\mu_d,\mu_s)=
~\frac{3}{\pi^2}~\sum_{i=u,d,s}~\int_0^{\sqrt{\mu_i^2-m_i^2}}~
dp~p^2(\sqrt{p^2+m_i^2}-\mu_i)+ B_{\rm eff} \, 
\label{omegakin}
\eeq 
where $\mu_u=\mu-\twothirds\mu_e$, $\mu_d=\mu_s=\mu+\third\mu_e$ are the
individual quark chemical potentials and $\mu$ and $\mu_e$ are the
baryon and electron chemical potentials respectively. We may neglect
the quark masses of the up and down quarks in quark matter since 
$m_{\rm light} \sim m_u \sim m_d \ll \mu$. On the other hand, the strange quark mass
is not negligible compared to $\mu$. In this work we will assume that
$ m_s < \mu$ so that an expansion in powers of $m_s/\mu$ is meaningful
and study three flavor quark matter. 

In neutral unpaired quark matter, the electron chemical potential
is determined by the condition of charge neutrality and is given by
\beq 
\mu_e = \frac{m_s^2}{4\mu} -
\frac{m_s^4}{48\mu^3} + \mathcal{O}\left(\frac{m_s^6}{\mu^5}\right)  \,,
\label{themue}
\eeq
Substituting this in Eq.~\ref{omegakin} and expanding in powers of
$m_s/\mu$ we obtain 
\bea
\label{omegams}
\Omega_{\rm unpaired}^{\rm neutral}(\mu) &=& -\frac{3}{4\,{\pi }^2}(1-c)\mu^4
 + \frac{3\mu^2\,m_s^2}{4\,\pi^2} 
 + \left(12\log\Bigl(\frac{m_s}{2\mu}\Bigr)-7\right)
   \frac{{{m_s}}^4}{32\,{\pi }^2}\nonumber \\
 &+& \frac{5 m_s^6}{576\pi^2\mu^2} + B_{\rm eff}
 + \mathcal{O}\left(\frac{m_s^8}{\mu^4}\right),
\eea 
where we have dropped terms of order $m_s^8/\mu^4$ and higher, and we
have introduced a parameter $c$ corresponding to 
the QCD-inspired corrections of Ref.~\cite{Fraga}, just
as we did in Sec.~\ref{sec:stage1}.
From Eq.~\eqn{omegams} we see that the expansion in powers of $m_s/\mu$ is
rapidly convergent, even for $m_s \sim \mu$.  The contribution to the free
energy from the $m_s^6/\mu^2$ term is less than one part in $10^4$
of the $\mu^4$ term
when $m_s\sim 300 $ MeV and $\mu=350$ MeV. This means that neutral unpaired
quark matter is really just a particular case of the phenomenological
parameterization explored in Sec.~\ref{sec:stage1}.
The FPS parameterization relies on the analysis of Freedman
and McLerran which is rigorous for massless quarks\cite{pertEoS}. We
therefore apply the effects of the ``perturbative'' QCD
correction only to the $\mu^4$ (\ie\ massless) part of the free energy.
Again, this is similar to the phenomenological
parameterization, where the QCD correction is a modification of the
coefficient of the $\mu^4$ term.

If we do not impose the neutrality condition, but expand in powers
of $\mu_e/\mu$ as well as $m_s/\mu$, 
we find that the expansion
does not converge nearly as well, so for charged unpaired quark matter
we must use the full form of the free energy \Eqn{omegakin}.
%

\subsection{Color-flavor-locked (CFL) quark matter}

In color-flavor locked matter, the pairing
locks the Fermi momenta of the all the quarks to a single value, 
requiring the number densities of up, down and strange quarks
to be equal \cite{cflneutrality}. This costs free energy, which is offset
by the pairing contribution
\beq \Omega_{\Delta} =
-\frac{3}{\pi^2}~\Delta^2\mu^2 + \mathcal{O}(\Delta^4) \,.
\eeq
Calculations of $\Delta$ with effective interactions yield values in
the range $10-100$ MeV \cite{colorSCreviews} for $\mu$ in the range
$300-600$ MeV. So it is reasonable to retain only the leading order
(in powers of $\Delta$) contribution. As in the case of the strange
quark mass, the $\Delta^4$ contribution has a weak (logarithmic)
dependence on $\mu$ and its contribution to the equation of state is
indistinguishable from $B_{\rm eff}$. 

When color and electric neutrality is imposed, there are no electrons
in the CFL phase since there are equal numbers of up, down, and strange
quarks, so the electron chemical potential $\mu_e=0$.
Expanding the free energy in powers of $m_s/\mu$, we find\cite{ARRW}
\bea
\Omega_{\rm CFL\ quarks}^{\rm neutral}&=&
 - \frac{3}{4\pi^2}(1-c)\mu^4
 + \frac{3 m_s^2 \mu^2}{4\pi^2} 
 - \frac{3\Delta^2\mu^2}{\pi^2} \nonumber \\
&+& \left(12\log\Bigl(\dsp\frac{m_s}{2\mu}\Bigr)-1\right)\frac{m_s^4}{32\pi^2} 
+ B_{\rm eff} +{\cal O}\left(\frac{m_s^6}{\mu^2}\right) \nonumber \\
&=& \Omega_{\rm unpaired}^{\rm neutral}
  + \frac{ 3 m_s^4 - 48\Delta^2\mu^2}{16\pi^2} 
  + \mathcal{O}\left(\frac{m_s^6}{\mu^2}\right)\ ,
\label{Omega_CFL_neutral}
\eea
where we have assumed that the unpaired and CFL phases have the same
bag constant.

We see from the first line of this expression that the
equation of state of neutral CFL matter is controlled by
the strange quark mass and pairing gap in a combination
that corresponds to the parameter $a_2$ in the phenomenological
parameterization \eqn{QM_EoS1}, if we identify
\beq
a_2 = m_s^2 - 4 \De^2\ .
\label{a2_CFL}
\eeq
Note that the $m_s^4$ terms do not follow such an identification,
but they are irrelevant because they are $\mu$-independent,
so they are a renormalization of the bag constant,
which we choose to give some specified transition density.
Therefore the neutral CFL equation of state, to the order that we have 
expanded it, is rigorously a function of the QCD-inspired correction $c$,
the chosen transition density $\rhocrit$, and $m_s^2 - 4 \De^2$.
However, the 
equation of state for neutral quark matter in general
is not rigorously a function of
$m_s^2 - 4 \De^2$ only, because 
for a given $m_s$ and $\De$ there is competition
between the neutral CFL phase and the neutral
unpaired phase, whose equation of state
is affected by $m_s$ alone. However, we expect the dependence
on the linearly independent variable $4 m_s^2 + \De^2$ to
be very weak. By independently varying $m_s$ and $\De$ we have 
verified that this is the case.
Of course, mixed phases are complicated by the fact that they
involve charged quark matter whose equation of state depends on
$\mu_e$ as an additional parameter.

%
%

We want to construct mixed phases so we must also know the
equation of state for charged CFL matter with $\mu_e\ne 0$. 
This has contributions from the quarks, the electrons, 
and from Goldstone bosons. The Goldstone bosons arise due to the 
spontaneous breaking of chiral symmetry, analogous to the mechanism
in vacuum \cite{CFL}. The quark contribution is independent
of $\mu_e$: CFL-paired quarks form an insulator with gap $\De$,
so as long as $\mu_e<\De$ there are no charged quasi-quark excitations,
so $Q = \p\Om/\p\mu_e = 0$.

The Goldstone bosons are a consequence of spontaneous breaking of 
chiral symmetry \cite{CFL,ARRW} and like the octet of pseudoscalar mesons 
of QCD in vacuum, can be described by an effective chiral field theory
\cite{effectivetheory}. When the electron chemical potential exceeds
the mass of the lightest negatively charged meson, which in the CFL
phase is the $\pi^-$, these mesons condense
\cite{kaons}. The free energy contribution of the meson condensate is
\begin{equation}
\Omega^{\rm GB}_{\rm CFL}(\mu,\mu_e)=-\frac{1}{2}f_{\pi}^2 \mu_e^2
 \left(1-\frac{m_{\pi}^2}{\mu_e^2} \right)^2 \ ,
\label{EoS_kaons}
\end{equation}
where the parameters are \cite{effectivetheory}
\beq
f_\pi^2 = \frac{(21 - 8 \ln 2)\mu^2}{36\pi^2}, \qquad
m_{\pi^-}^2 = \frac{3 \De^2}{\pi^2f_\pi^2 } m_s(m_u+m_d)\,.
\label{pion_params}
\eeq 
We used $m_u=3.75~\MeV$, $m_d=7.5~\MeV$.
Finally, the free energy contribution from electrons and muons is
given by 
\beq 
\Om^{\rm leptons}(\mu_e) =
\sum_{i=e^-,\mu^-}
\frac{1}{\pi^2} ~\int_0^{\sqrt{\mu_e^2-m_i^2}}
dp~p^2(\sqrt{p^2+m_i^2}-\mu_e) \,.  
\label{EoS_laptons}
\eeq 
%
The total free energy for CFL quark matter is then
\beq
\Omega_{\rm CFL}^{\rm charged}(\mu,\mu_e) =
\Omega_{\rm CFL\ quarks}^{\rm neutral}(\mu)
+ \Omega^{\rm GB}_{\rm CFL}(\mu,\mu_e)
+ \Omega^{\rm leptons}(\mu_e)
\eeq

\subsection{A bag-model hybrid star that masquerades as a neutron star.}

\begin{figure}[tb]
\begin{center}
 \includegraphics[width=0.8\textwidth,angle=0]{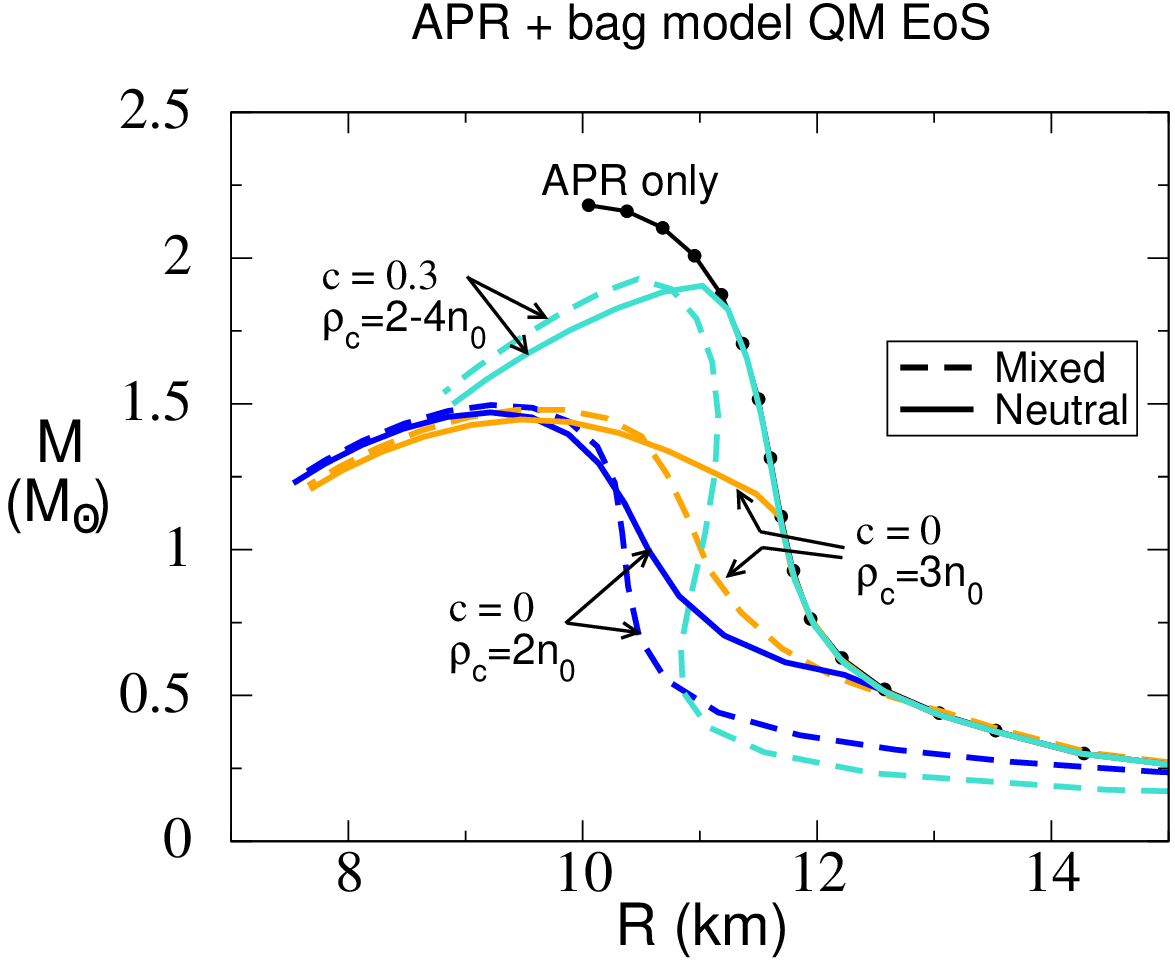}
\end{center}
\caption{\small
$M(R)$ relationship for hybrid stars involving quark matter obeying
the simple bag model equation of state, with various values of $c$
and $\rhocrit$, and $m_s=180~\MeV$ and $\De=0$.
The line with black dots is the $M(R)$ relation for a pure nuclear
APR star. The solid lines are for hybrid stars with 
homogeneous neutral APR and CFL phases, the
dashed lines are for hybrid stars with mixed APR+CFL phases.
We see that the maximum mass $\Mmax$ is very sensitive to the perturbative 
correction $c$, but not to the transition density, or the occurrence of
mixed phases. For $c=0.3$, which is close to the value suggested
in \cite{Fraga}, $\Mmax\approx 1.9~\Msolar$.
}
\label{fig:stage2_heavystar}
\end{figure}

Using the bag model equations of state for unpaired and CFL 
quark matter, we can proceed as in Sec.~\ref{sec:stage1}
to solve the TOV equation and obtain $M(R)$ curves for
hybrid stars. We first show a specific example, in
Fig.~\ref{fig:stage2_heavystar}. We have chosen the bag model
parameters to give the same equation of state for neutral
matter as was studied
using the phenomenological parameterization in 
Fig.~\ref{fig:stage1_heavystar}. Now, however, we can also
study mixed phases. 
The solid lines, for homogeneous neutral quark and nuclear matter phases,
are the same as in Fig.~\ref{fig:stage1_heavystar}. The dashed
lines show the $M(R)$ relation when mixed phases are allowed.
We see that although the overall shape of the
$M(R)$ curve is quite different when mixed phases
are present the maximum mass is not significantly affected.
The maximum mass of a compact star is therefore insensitive to 
the surface tension of the interface
between quark matter and nuclear matter. 
This is because the maximum mass
configuration is characterized by a baryon density that is large
compared to the transition density and most of the star is either in
the homogeneous quark matter phase, or in a mixed phase that is
dominated by quark matter.

We can also see that, as in
the phenomenological model explored in Sec.~\ref{sec:stage1},
the maximum mass
is very sensitive to the QCD correction $c$, and
relatively insensitive to the transition density.
All the curves in Fig.~\ref{fig:stage2_heavystar} are for 
$m_s=180~\MeV, \De=0$, so from this figure
we cannot judge the sensitivity to those
parameters.

\subsection{Maximum mass as a function of bag model parameters}

\begin{figure}[tb]
\begin{center}
\includegraphics[width=0.8\textwidth,angle=0]{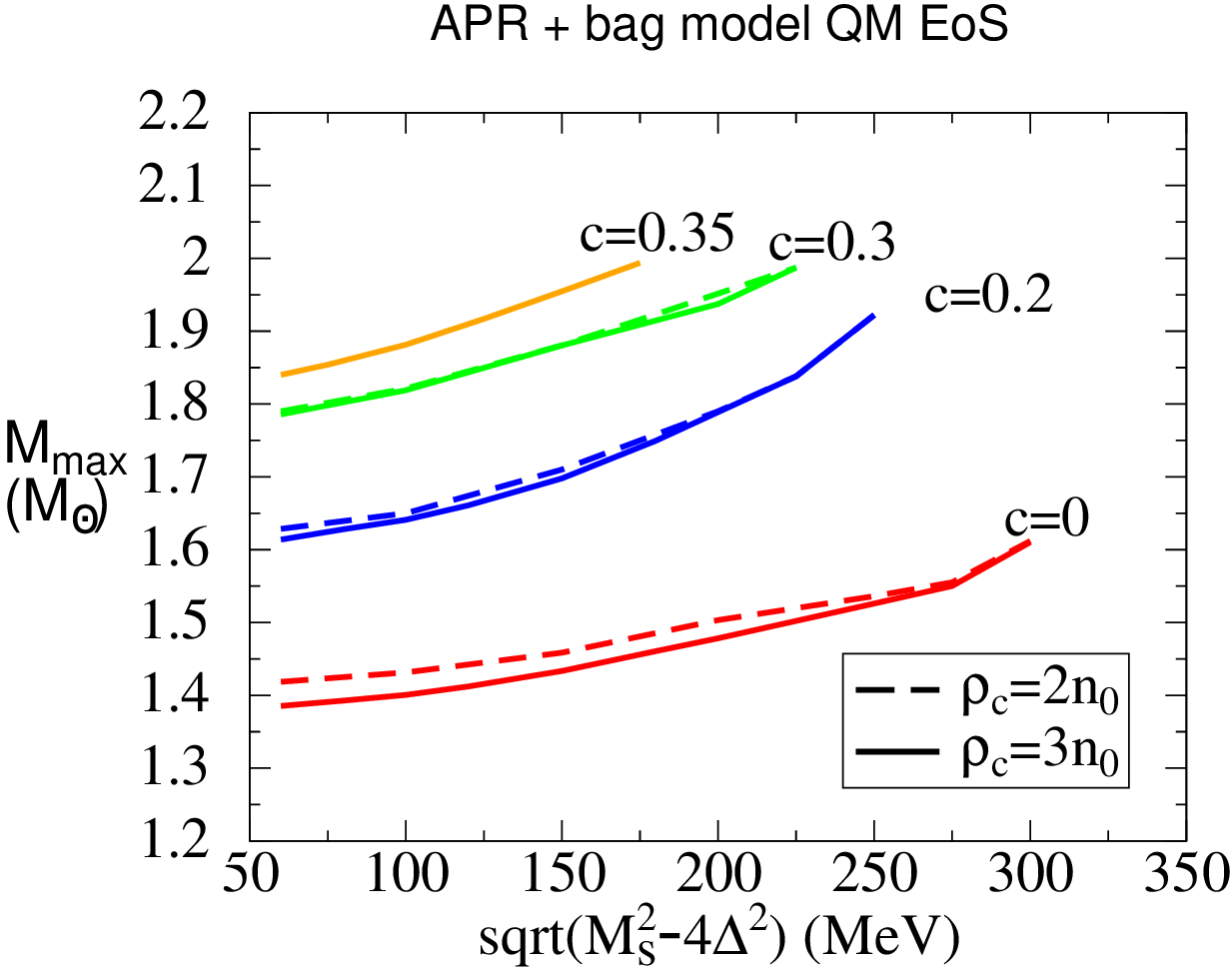}
\end{center}
\caption{\small
Maximum mass $\Mmax$ for hybrid stars containing quark matter obeying
the simple bag model equation of state. We show the dependence on
$\sqrt{m_s^2-4\De^2}$.
Increasing this quantity decreases 
$B_{\rm eff}$ at fixed $\rhocrit$, giving rise to larger maximum masses. 
The maximum mass is only weakly dependent on the transition density,
particularly when the perturbative correction is included.
}
\label{fig:stage2_vary_y}
\end{figure}

In Fig.~\ref{fig:stage2_vary_y} we give a more complete picture, by plotting
the maximum mass of the star as a function of $\sqrt{m_s^2-4\De^2}$.
Varying the independent variable,
$\sqrt{4m_s^2+\De^2}$, has negligible effect, as one would expect.
We plot this relationship
for $\rhocrit=3\,\nsat$ and $\rhocrit=2\,\nsat$. The stars
with small $m_s^2-4\De^2$ correspond to hybrid stars containing CFL quark
matter while those with large $m_s^2-4\De^2$ correspond to stars with unpaired
quark matter. We see that the largest masses are obtained by turning
$m_s^2-4\De^2$ up to a large value, corresponding to a strange quark mass 
in the range $200~\MeV$ to $300~\MeV$. This is the same situation that was 
described in Sec.~{\ref{sec:maxM} since the effective 
bag constant must decrease with increasing $m_s^2-4\De^2$ at fixed
$\rhocrit$. As mentioned earlier, stars dominated by quark matter have
$\Mmax \sim 1/\sqrt{B_{\rm eff}}$\cite{Witten:1984rs}.
We also see that the value of the transition density is not very
important: $\Mmax$ is only weakly dependent on the transition density,
particularly when the perturbative correction is included.

%

\section{Conclusions} 

We have studied the mass-radius relationship for hybrid compact stars,
with a nuclear matter crust (described by the APR equation of state)
and a quark matter core. We used a phenomenological parameterization
\eqn{QM_EoS} of the neutral quark matter equation of state, and
also a simple MIT bag model that allowed us to construct mixed phases.
In both cases we included
a QCD correction parameter $c$ in the quark matter equation
of state, and we found that increasing its value
from zero (no QCD corrections) to reasonable value ($c=0.35$
\cite{Fraga}) increases the maximum hybrid star mass from about
$1.6~\Msolar$ to about $2~\Msolar$. 
This is clear from Fig.~\ref{fig:stage1_vary_y} (for the
phenomenological parameterization) and Fig.~\ref{fig:stage2_vary_y}
(for the bag model).
The reason is that increasing $c$ hardens the quark
matter equation of state, making it almost indistinguishable from the
APR nuclear equation of state (see Fig.~\ref{fig:EoS_fixrho}).
It is important to note that we achieve these masses with
reasonable transition densities $\rhocrit$ of order 2 to 3 times $\nsat$.
Our stars have a proper crust of nuclear matter:
we are not resorting to low transition densities that yield ``hybrid''
stars that are actually quark stars with a tiny shell of nuclear matter
around the outside.

We conclude that it is harder than previously thought for a simple
mass measurement to rule out the presence of quark matter in compact
stars. Currently published 
measurements of the masses of compact stars
are all consistent with a maximum mass of $\Mmax\approx
1.7~\Msolar$ at the $2\si$ level.  Our results show that maximum
masses of up to $2~\Msolar$ can be accommodated by models of hybrid
stars with reasonable quark matter equations of state.

We also note that our $M(R)$ curves are consistent with the
constraint obtained from measurements of red shifts of Iron absorption
lines in the low-mass X-ray binary EXO0748-676 \cite{Cottam}. If that
constraint were plotted in our Fig.~\ref{fig:stage1_heavystar} it would
intersect our $c=0.3$ curve at $M\approx 1.7~\Msolar$.

This naturally raises two related questions: what sort of mass
or radius observation {\em would} provide evidence against the
presence of quark matter in neutron stars? And what sort of observation would
provide evidence for the presence of 
color-superconducting quark matter in particular?
 
Obviously a mass measurement above $2\Msolar$ would give reason to
doubt the presence of quark matter. Other than that, it seems
difficult to diagnose the presence of quark matter via $M(R)$
measurements. The regions of $M(R)$ space that can be reached by
hybrid quark-nuclear stars are the same as those that can be reached
by hadronic matter stars, once moderately exotic phenomena such as 
kaon condensation or
hyperon production are allowed \cite{Lattimer:2000nx}.  
Observation of an
object with a small radius ($R\approx 7$ to $10~\km$ at $M\approx
1.4~\Msolar$) would rule out simple nucleonic matter, but would not favor
quark matter over the exotic forms of hadronic matter. There are
regions of parameter space (very small $M$ and $R$, for example)
that can only be reached by pure quark matter objects, which
only exist if quark matter is absolutely stable ($\rhocrit=0$).
Other regions of the parameter space are
inconsistent with both hadronic and quark matter.

Demonstrating the presence of {\em color-superconducting} quark matter
via $M(R)$ measurements appears to us to be very difficult.
However, this is not for the naive reason that ``the color
superconductivity contribution to the pressure is suppressed by ${\cal
O}(\De^2/\mu^2)$''.  As previously noted \cite{Lugones:2002ak,AlfRed},
the leading $\mu^4$ contribution is mostly canceled by the bag
constant, so the subleading $\De^2\mu^2$ term is potentially
important. However, in practice it is not detectable. Firstly, the
bag model
equation of state depends on the color superconducting gap $\De$ via
the linear combination $m_s^2-4\De^2$, so an accurate
determination of $m_s$ at high density would be needed to expose the
presence of a nonzero $\De$.  Secondly, the dependence of the $M(R)$
relation on $m_s^2-4\De^2$ is not particularly strong 
(see Fig.~\ref{fig:stage2_vary_y})
when a physical parameter $\rhocrit$ is assumed known, rather than the bag
constant.

If color-superconducting quark matter is to be found in compact stars,
it seems more likely that it will be detected via its effects on
transport properties. Color superconductivity drastically alters
these, and possible signatures are being actively investigated. These
include cooling (sensitive to heat capacity and neutrino emissivity
and opacity) \cite{neutrinos}, $r$-mode spin-down (sensitive to bulk
and shear viscosity) \cite{rmode}, and glitches (sensitive to
superfluidity and rigid structures) \cite{glitch}. 

\vspace{3ex}
{\samepage 
\begin{center} {\bf Acknowledgments} \end{center}
\nopagebreak
We have had useful conversations with J. Lattimer and M. Prakash. 
The research of M.A. and M.B. is supported by the Department of Energy
under grant number DE-FG02-91ER40628. 
The research of S.R. is supported by the Department of Energy under
contract W-7405-ENG-36.  
The research of M.P. is supported by DOE contract DE-AC05-84ER40150
under which the Southeastern Universities Reseach Association operates
the Thomas Jefferson National Accelerator Facility.
}

\end{document}